\documentstyle{mn}
\input epsf
\begin{document}
\journal{astro-ph/0101149}
\title[Reionization and the COBE normalization]{The effect of reionization on 
the COBE normalization}
\author[L.~M.~Griffiths and A.~R.~Liddle]{Louise M.~Griffiths$^1$ and Andrew 
R.~Liddle$^2$\\
$^1$Astrophysics, Nuclear and Astrophysics Laboratory, Keble Road, Oxford
OX1 3RH, United Kingdom.\\
$^2$Astronomy Centre, University of Sussex, Falmer, Brighton BN1 9QJ, United 
Kingdom.}
\maketitle
\begin{abstract}
We point out that the effect of reionization on the microwave anisotropy power 
spectrum is not necessarily negligible on the scales probed by COBE. It can lead 
to an upward shift of the COBE normalization by more than the one-sigma error 
quoted ignoring reionization. We provide a fitting function to incorporate 
reionization into the normalization of the matter power spectrum.
\end{abstract}
%%%%%%%%%%%%%%%%%%%%%%%%%%%%%%%%%%%%%%%%%%%%%%%%%%%%%%%%%%%%%%%%%%%%%%

\section{Introduction}

One of the most important uses of the COBE observations of large-angle
cosmic microwave background anisotropies (Smoot et al.~1992; Bennett
et al.~1996) is to normalize the power spectrum of matter fluctuations
in the Universe. Accordingly, several papers have been written quoting
fitting functions for this normalization as a function of various
cosmological parameters. Because COBE only probes scales larger than
the horizon at last scattering, it is insensitive to parameters
governing physical processes within the horizon, such as the Hubble
parameter $h$ and the baryon density $\Omega_{{\rm B}}$. It does
however depend on the matter density $\Omega_0$, the cosmological
constant density $\Omega_\Lambda$, the spectral index (tilt) of
density perturbations $n$ and the presence of gravitational waves
(usually parametrized by a quantity $r$). Existing literature has
provided fitting functions for spatially-flat models and open models
including tilt (Bunn \& White 1997), and for spatially-flat models
with both tilt and gravitational waves (Bunn, Liddle \& White 1996).

There is however one other parameter which can significantly alter the
normalization, which is the optical depth $\tau$ for rescattering of
microwave photons at lower redshift. The absence of absorption by
neutral hydrogen in quasar spectra, the Gunn--Peterson effect (Gunn \&
Peterson 1965; see also Steidel \& Sargent 1987; Webb 1992), tells us
that the Universe must have reached a high state of ionization by the
redshift of the most distant known quasars, around five. Scattering
from the created free electrons predominantly has the effect of
damping out the primary anisotropies, and the requirement that the
observed peak at $\ell \sim 200$ is not destroyed sets an upper limit
on the optical depth. Griffiths, Barbosa \&
Liddle (1999) obtained a limit of around $\tau \la 0.4$ for the most
plausible cosmological parameters; including new CMB data from BOOMERanG-98 and 
MAXIMA-1, plus other non-CMB constraints, can strengthen this somewhat (Tegmark, 
Zaldarriaga \& Hamilton 2000). In
addition to the damping, the rescattering generates a modest amount of new 
anisotropy
via the Doppler effect.

\begin{figure}
\centering 
\leavevmode\epsfysize=6cm \epsfbox{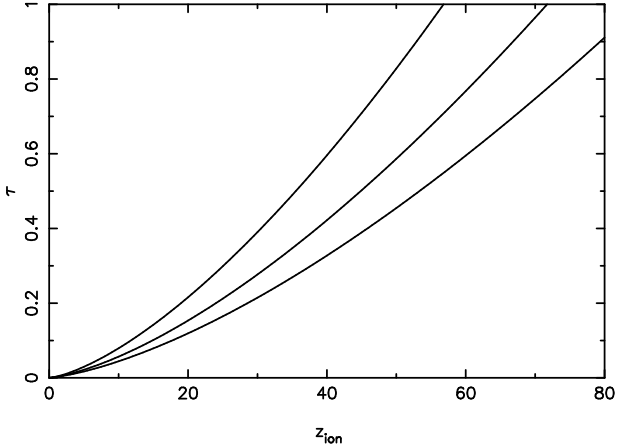}\\ 
\caption[tau_z]{\label{tau_z} Optical depth for instantaneous
reionization at redshift $z_{{\rm ion}}$. From top to bottom the
curves are $\Omega_0 = 0.3$, $0.6$ and $1$. We took $\Omega_{{\rm B}}
h^2 = 0.02$ and $h=0.65$.}
\end{figure}

Several mechanisms for reionization, which requires a source of
ultra-violet photons, have been discussed, and are extensively
reviewed by Haiman \& Knox (1999).  In the two most popular models,
the sources are massive stars in the first generation of galaxies, or
early generations of quasars. Calculations are sufficiently uncertain
as to give no clear guidance as to where within the currently allowed
range reionization might have occurred. Assuming spatial flatness and
instantaneous full reionization at redshift $z_{{\rm ion}}$, the optical
depth is related to the redshift of reionization by (e.g.~Griffiths et
al.~1999)
\begin{equation}
\label{e:tauz2}
\tau(z_{{\rm ion}}) = \frac{2\tau^* }{3\Omega_0} \, \left[ \left(1-\Omega_0
        + \Omega_0 (1+z_{{\rm ion}})^3 \right)^{1/2} - 1 \right] \,,
\end{equation}
where 
\begin{equation}
\tau^* = \frac{3 c H_0 \, \Omega_{{\rm B}} \, 
        \sigma_{{\rm T}}}{8\pi G m_{{\rm p}}} \times 0.88 
        \simeq 0.061 \, \Omega_{{\rm B}} h \,.
\end{equation}
Sample curves are shown in Figure~\ref{tau_z}. We should expect $\tau$
to lie anywhere between about 0.02 and 0.4.

We note that this does not include the contribution to the optical depth of the 
residual ionization left over after recombination, which for typical 
cosmological parameters is about 0.001 and which can contribute a further 
optical depth of a few percent between $z_{{\rm ion}}$ and $z_{{\rm rec}}$ 
(Seager, Sasselov \& Scott 2000).

Because the rescattering happens at low redshifts, it can affect the
microwave anisotropies on much larger angular scales than can causal
processes at last scattering.  The normalization of the power spectrum
from COBE is primarily from multipoles with $\ell \sim 10$, and
because the normalization is so accurate, it turns out that reionization can 
lead to a significant
effect. We shall concentrate on the fitting functions quoted by Bunn
\& White (1997) and Bunn et al.~(1996), whose quoted error on the
dispersion, $\delta_{{\rm H}}$, is 7 per cent statistical, with fit
and other systematics bringing this up to 9 per cent. This is
equivalent to a shift in the radiation angular power spectrum of just
under 20 per cent.

\section{Correcting the normalization for reionization}

We consider only spatially-flat cosmologies, as significantly open
models are now excluded (Jaffe et al.~2000). The COBE normalizations are readily 
obtained
from the publically-available {\sc cmbfast} (Seljak \& Zaldarriaga 1996) and 
{\sc camb} programs (Lewis, Challinor \& Lasenby
2000). Figure~2 shows a series of curves with varying optical depth,
where the normalization of the matter power spectrum has been kept
fixed, focusing on the region relevant to the COBE observations. The other 
cosmological parameters are those of the favoured low-density flat model with 
$\Omega_0 = 0.3$, $n=1$,
$h = 0.65$, $\Omega_{{\rm B}} h^2 = 0.02$ and no
gravitational waves. To a first approximation, normalizing to COBE
will shift the curves to the same amplitude at $\ell \simeq 10$. We
use the power spectrum normalizations output from the code, which are
computed by fitting a quadratic to the $C_\ell$ spectrum and
implementing a fitting function from Bunn \& White (1997).

\begin{figure}
\centering 
\leavevmode\epsfysize=5.3cm \epsfbox{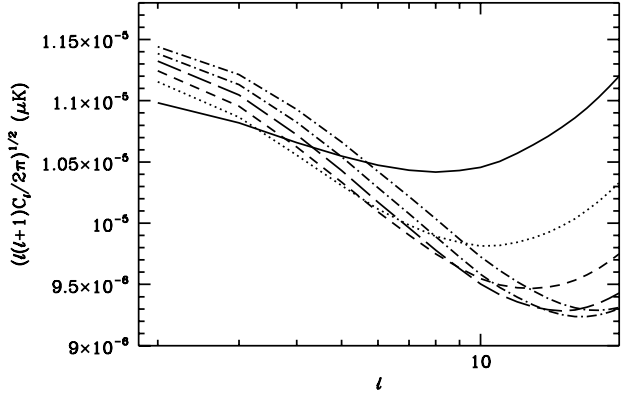}\\ 
\caption[cls]{\label{cls} A set of $C_\ell$ curves for the standard
cosmology, showing (from bottom to top on the left-hand edge) optical depths 
$\tau = 0$, $0.1$, $0.2$, $0.3$, $0.4$ and $0.5$.}
\end{figure}

Figure~2 shows two separate physical effects in operation. On the right-hand 
side of the plot we mainly observe the effect of reionization damping erasing 
the initial anisotropies, a process described in detail by Hu \& White 
(1997); there is also some regeneration of anisotropies on those scales from the 
Doppler effect in scattering from the reionized electrons. Bearing in mind that 
the COBE normalization is particularly sensitive to multipoles around $\ell = 
10$, for low $\tau$ the damping is the dominant effect in altering the power 
spectrum normalization.

A more interesting effect is the rise of the multipoles, including the lowest, 
as the optical depth is increased, an effect which reaches 10 per cent for $C_2$ 
at 
$\tau = 0.5$. The effect has a subtle origin. There are contributions to the 
anisotropies both from the original last-scattering surface and from the 
reionization scattering surface. Between recombination and reionization 
large-scale perturbations (i.e.~those with comoving wavenumbers much less than 
the Hubble scale) cannot evolve, but smaller scale ones can. A given $C_\ell$ 
actually receives contributions from quite a wide range of comoving wavenumbers, 
so that the low multipoles exhibit 
some sensitivity to what is happening on smaller scales and this results in the 
increase in power.\footnote{We are 
indebted to Matias Zaldarriaga for correspondence on this issue.} Although the 
effect is not large, it is significant at the level of the COBE normalization, 
and once $\tau$ exceeds around 0.3 the rise in the radiation power spectrum from 
newly-generated anisotropies actually becomes more important than  the effect of 
reionization damping. 

In Figure~3, we plot the change in the COBE normalization of the
matter power spectrum as a function of $\tau$ for various choices of
$\Omega_0$. The quantity $\delta_{{\rm H}}$, defined as in Bunn et al.~(1996) 
and Bunn \& White (1997), measures the dispersion of the matter distribution, 
and the COBE normalization of it has a statistical error of 7 per cent. The 
power 
spectrum normalization goes as $\delta_{{\rm H}}^2$. We see that the 
reionization has a significant effect on the normalization, corresponding to 
roughly a one-sigma shift for a wide range of optical depths. For low optical 
depths the reionization damping dominates and the normalization increases, 
reaching a maximum at $\tau \simeq 0.3$. As $\tau$ increases further the 
reionization 
damping moves to smaller angular scales and becomes less significant at $\ell 
\simeq 10$ than the 
regenerated anisotropies, and the normalization begins to fall.

As seen in Figure~3, there is a weak dependence on $\Omega_0$, and indeed there 
are similar dependences on the parameters $h$ and $\Omega_{{\rm B}}$. The 
dependences arise because these parameters alter the reionization redshift, and 
hence characteristic angular size, corresponding to a given optical depth. The 
effect of varying these 
parameters is typically at the one or two percent level, hence much smaller than 
the effect of the optical depth.

\begin{figure}
\centering 
\leavevmode\epsfysize=5.6cm \epsfbox{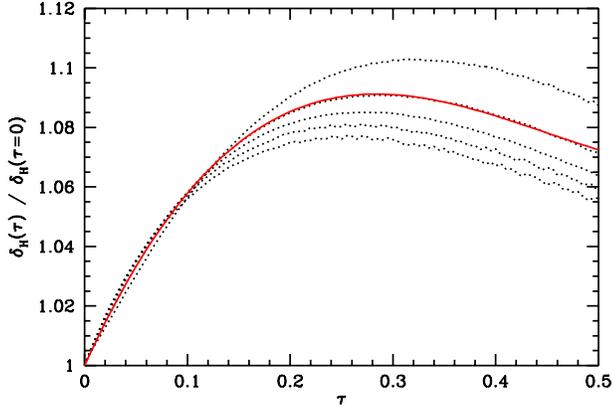}\\ 
\caption[norm]{\label{norm} The COBE normalization as a function of
$\tau$, is shown by the dashed lines for (from top to bottom) $\Omega_0 = 0.2$, 
$0.4$, $0.6$, $0.8$ and $1.0$. The solid line shows the fit quoted in equation 
(\ref{e:fit}).}
\end{figure}

There is no point in trying to fit the dependence of the reionization correction 
on $h$ and $\Omega_{{\rm B}}$, since published normalizations ignore the effect 
of these parameters in the case with no reionization as it is well within the 
statistical error from cosmic variance. Equally, although quoted 
fitting functions do give a dependence on $\Omega_0$, the additional $\Omega_0$ 
dependence of the reionization correction is at the same level (one or two 
percent) as 
those ignored effects, and so there is no incentive to try to include it 
either. Therefore, to sufficient accuracy one can ignore the dependence of the 
normalization on parameters other than $\tau$, and the correction to the 
normalization can then be expressed via a $\tau$-dependent fitting function. We 
chose to fit for $\Omega_0 = 0.4$ as it lies roughly centrally amongst the 
models we studied and is close to 
the currently favoured value. A good fit is given by the form
\begin{equation}
\label{e:fit}
\frac{\delta_{{\rm H}}(\tau)}{\delta_{{\rm H}}(\tau = 0)} = 1 + 0.76 \tau - 1.96 
\tau^2 + 1.46 \tau^3 \,,
\end{equation}
as shown in Figure~3, which is reliable up to $\tau$ of 0.5. This correction can 
be applied to equation (29) in Bunn 
\& White
(1997) and to equations (15), (A4) and (A5) in Bunn et al.~(1996); note that 
these fitting functions have quoted fit errors of up to 3 per cent though they 
are 
usually within 1 per cent. Even allowing for possible variation of 
other parameters including $\Omega_0$, the fit error for our correction is 
within 2 per cent which, given 
the error in the COBE normalization, should be more than adequate for the 
foreseeable future.

\section{Summary}

We have quantified the effect of reionization on the COBE
normalization of the matter power spectrum. For values of the optical
depth in the centre of the currently-allowed region, reionization
leads to a significant enhancement of the COBE-normalized matter power spectrum, 
which should be accounted for in attempts to constrain cosmological
parameters by combining other data sets with the COBE normalization. [The effect 
is of course automatically included in analyses which simultaneously fit cosmic 
microwave background data and other data, except that most such analyses have 
not 
so far included reionization, an exception being Tegmark et al.~(2000).] 

We have provided a simple fitting function which allows this correction to be 
incorporated into published fitting functions.

%%%%%%%%%%%%%%%%%%%%%%%%%%%%%%%%%%%%%%%%%%%%%%%%%%%%%%%%%%%%%%%%%%%%%%%%
\section*{ACKNOWLEDGMENTS}

LMG was supported by PPARC.  We thank Antony Lewis and Matias
Zaldarriaga for clarifying discrepancies between the {\sc camb} and
{\sc cmbfast} codes, and Matias Zaldarriaga for providing a patch. We further 
thank Matias Zaldarriaga for important correspondence on the low $\ell$ 
behaviour in reionized models. 

%%%%%%%%%%%%%%%%%%%%%%%%%%%%%%%%%%%%%%%%%%%%%%%%%%%%%%%%%%%%%%%%%%%%%%%%

%%%%%%%%%%%%%%%%%%%%%%%%%%%%%%%%%%%%%%%%%%%%%%%%%%%%%%%%%%%%%%%%%%%%%%

\bsp
\end{document}